\documentclass[authoryear,preprint,review,3p]{elsarticle}

\usepackage{tikz}
\usepackage{array}
\usepackage{chemformula}
\usepackage{multirow}
\usepackage{siunitx}
\usepackage{color}
\journal{Journal of Marine and Petroleum Geology}

\begin{document}
\begin{frontmatter}

\title{Dynamic in-situ imaging of methane hydrate formation and self-preservation}
\author[lund]{Viktor~V.~Nikitin\corref{cor1}}
\ead{viktor.nikitin@maxiv.lu.se}

\author[ipgg]{Geser~A.~Dugarov}
\author[ipgg,nsu]{Anton~A.~Duchkov}
\author[ipgg]{Mikhail~I.~Fokin}
\author[ipgg]{Arkady~N.~Drobchik}
\author[aps]{Pavel~D.~Shevchenko}
\author[aps]{Francesco~De~Carlo}
\author[lund]{Rajmund~Mokso}

\cortext[cor1]{Corresponding author}
\address[lund]{Max IV Laboratory, Lund University,  2 Fotongatan, 22484 Lund,~Sweden}
\address[ipgg]{Institute of Petroleum Geology and Geophysics SB RAS,3 Ac. Koptyuga ave., 630090 Novosibirsk,~Russia}
\address[nsu]{Novosibirsk State University, 1 St. Pirogova, 630090 Novosibirsk,~Russia}
\address[aps]{Advanced Photon Source, Argonne National Laboratory, 9700 S. Cass Avenue, 60439 Lemont, IL, USA}

\begin{abstract}
% 	\begin{linenumbers}
    
    We present the results of dynamic in-situ 3D X-ray imaging of methane hydrates microstructure during methane hydrate formation and decomposition in sand samples. Short scanning times and high resolution provided by synchrotron X-rays allowed for better understanding of water movement and different types of gas-hydrate formation. Complementing previous observations, we conclude that the process of gas-hydrate formation is accompanied by the water movements caused by cryogenic water suction that happens in short sequences with longer equilibrium states in between (when the water is immobile). Based on the 3D microstructure we identified two types of gas-hydrate formation: (i) into the gas pockets and (ii) inside water volumes. During the decomposition in the self-preservation mode (pressure drop at negative temperatures) the latter remains more stable compared to the hydrate formed as growing into the gas pocket. This means that the history of the gas-hydrate formation influences its behaviour at the decomposition stage (e.g. gas-hydrate production).
    
    % This paper reports results of the synchrotron-based tomography experiments for studying the methane-hydrate formation/decomposition in sand samples. 
    % Fast scanning times (70 secs) provided successive images (every 2-15 min) for a detailed study of these processes. 
    % During the methane gas-hydrate formation we observed water flows that we connect to the effect of cryogenic suction due to the hydrate formation. These appeared to be multiple fast water flows in pore space followed by periods of the water position stability. 
    % Dynamic images allow identifying two types of gas-hydrate formation: into the gas pockets and inside water volumes. 
    % These two gas-hydrate types show different properties during the decomposition in the self-preservation mode - pressure drop at negative temperatures. 
    % Gas hydrate formed in water volumes remains more stable compared to hydrate formed as growing into the gas pocket.
    % This means that the history of the gas-hydrate formation influences its behaviour at the decomposition stage (e.g gas-hydrate production). 
% 	\end{linenumbers}
\end{abstract}

% \begin{graphicalabstract}
% \includegraphics{figs/fig5.pdf}
% \end{graphicalabstract}

% \begin{highlights}
% \item Two types of hydrate growth mechanisms were observed: into gas pockets and inside water volumes.
% \item Only hydrates formed in water volumes were self-preserved at negative temperatures after dropping the pressure.
% \item Multiple fast water flows were captured during hydrate formation.
% \end{highlights}


\end{frontmatter}

\section{Introduction}

Natural gas hydrates are solid crystals that are mostly formed from water and methane. They accumulate worldwide in favorable thermobaric conditions (in permafrost regions on land and in bottom sediments offshore) and form hydrocarbon resources exceeding those in traditional and shale deposits \citep{birchwood2010developments}. They may also affect humanity in various other ways including flow assurance, safety issues, possible impact on ecology and climate change \citep{koh2007natural}. 
There are different ways of learning more about natural gas-hydrate systems: studying natural pressurized hydrate-bearing cores \citep{yoneda2017pressure,jin2016situ}, geophysical characterization of natural gas-hydrate accumulations \citep{riedel2010geophysical}, pilot experiments on the gas-hydrate production \citep{white2009designing,sun2014production}. 

For a better understanding of natural gas-hydrate systems we need a better knowledge of the hydrate formation/decomposition and associated processes in realistic systems. We are also interested in studying how these processes affect the rock  physical properties. This is important for further development of the geophysical methods for gas-hydrate accumulations exploration and monitoring. In most cases this is done by using specialized laboratory setups for forming synthetic gas hydrates in rock samples and studying their physical properties \citep{waite2009physical}. It turns out that macro-properties highly depend on the sample microstructure including matrix composition, pore structure, hydrate morphology, etc.   

So it is clearly beneficial to combine the laboratory experiments on forming and studying hydrate-bearing samples  with their imaging at different scales. For example, different scenarios of hydrate formation result in different hydrate morphology in pore filling, which considerably affects acoustic macro-properties  \citep{waite2009physical,priest2009influence,dugarov2019models}: non-cementing (pore-filling) hydrate formation does not affect acoustic velocities much, while cementing hydrate formation type results in a faster increase of acoustic velocities. 
X-ray Computed Tomography (CT) is widely used to image the detailed structure of rocks which have a complicated porous structure consisting of materials very different in properties: mineral particles, gas, multi-phase liquids. 
Standard laboratory X-ray CT devices require long exposure time (typically from half an hour to several hours) for imaging with micrometer resolution. They work well for imaging the microstructure of a static sample or follow very slow processes. microstructure of the methane-hydrate formation was reported in \citep{zhao2015microstructural} with some conclusions on distribution of the gas-hydrate particles and the matrix grains. 
\cite{lei2018strategies} showed that the gas-hydrate formation is displacive and segregated since the hydrate extracts water from the sediments. 
Later \cite{lei2019pore} reported evidence of water movement and described 3 types of hydrate formation mechanisms: hydrate growth over sand particles, hydrate growth by water invasion into gas pockets (via hydrate tubes), and poet-formation hydrate morphology evolution (via diffusion of water vapor or dissolved methane).
%in excess-water environments. 
%\textcolor{green}{Long scanning times cause difficulties in studying time dynamics of these processes.} 

During the last decade, X-ray imaging has developed into a useful tool to visualize in-situ processes in geomaterials. At the resolution level of \SI{1}{\micro\meter} for parallel X-ray geometry \citep{fusseis2014,saif2016}, and of \SI{20}{\nano\meter} for transmission X-ray microscope geometry \citep{deandrade2016}, synchrotron-based tomography is outperforming all other three-dimensional imaging methods for fast dynamic processes in bulky samples \citep{gibbs2015}. Characteristic time for one full scan is about 1-5~min for microtomography and about 5-\SI{60}{\minute} for nanotomography, depending on the sample size, instrument optics, and hardware limitations. Recent instrument developments \citep{Mokso2017} allow to capture processes with sub-second temporal resolution.
Important applications of fast imaging include multi-phase fluid flow in porous rocks \citep{Youssef2013,blunt2013pore,fusseis2014low}, deformation and geomechanical testing of samples \citep{BakerNComm2012,li2015dynamic,wang2016deformation}, and hydraulic fracturing \citep{kiss2015synchrotron}. 

%\textcolor{green}{Studying gas-hydrates is a relatively new application that can greatly benefit from 4D X-ray imaging. }
Natural gas-hydrate systems are specifically rich in dynamic chemical-physical processing with a wide range of time scales. Therefore dynamic tomographic imaging can contribute substantially to the understanding of these systems. Just to mention some of them: gas-hydrate formation takes hours to days in the laboratory experiments, gas-hydrate decomposition takes several seconds (or minutes to hours in the self-conservation mode), water/gas redistribution during gas-hydrate formation/decomposition may take seconds to minutes, water freezing/thawing cycles take minutes, etc. 

In our experiments we aim at working with the methane hydrate in order to mimic natural gas-hydrate systems while increasing resolution and decreasing the scanning time to image fast processes.
Use of synchrotron-based micro-CT imaging allows for fast scanning rate (up to \SI{70}{\second}) compared to laboratory X-ray sources requiring several hours to measure high-resolution datasets \citep{chen2018fastXe, lei2019pore}. At the same time, a high-intense coherent beam yields improvement in reconstruction quality with employing phase retrieval procedures. In \citep{chaouachi2015microstructural} and \citep{yang2016synchrotron} the authors used synchrotron X-ray micro-CT for imaging formation and dissociation of Xenon hydrate which ensures good contrast between gas-hydrate and water phases. However, Xenon and methane gas hydrates may have different properties and differ in formation/decomposition processes as well  \citep{lei2018salinityForCT,haynes2014}.  In this work, we operate with the methane gas and salty water in order to model the close to reality process.

The rest of the paper is organized as follows. In Section 2 we present the setup of the tomography experiment, a scheme for the data acquisition process, and reconstruction procedures. Section 3 shows main results and observations made from analyzing the reconstructed images. In Section 4 we summarize the results and present outlook for our next tomographic experiments for studying gas hydrates. 

%%%%%%%%%%%%%%%%%%%%%%%%%%%%%%%%%%%%%%%%%%%%%%%%%%%%%%%%%%%%%%%%%%%%%%%%%%%%%%
\section{Experiment Setup and Data Acquisition}

\begin{figure*}
  \centering
  \includegraphics[width=0.95\textwidth]{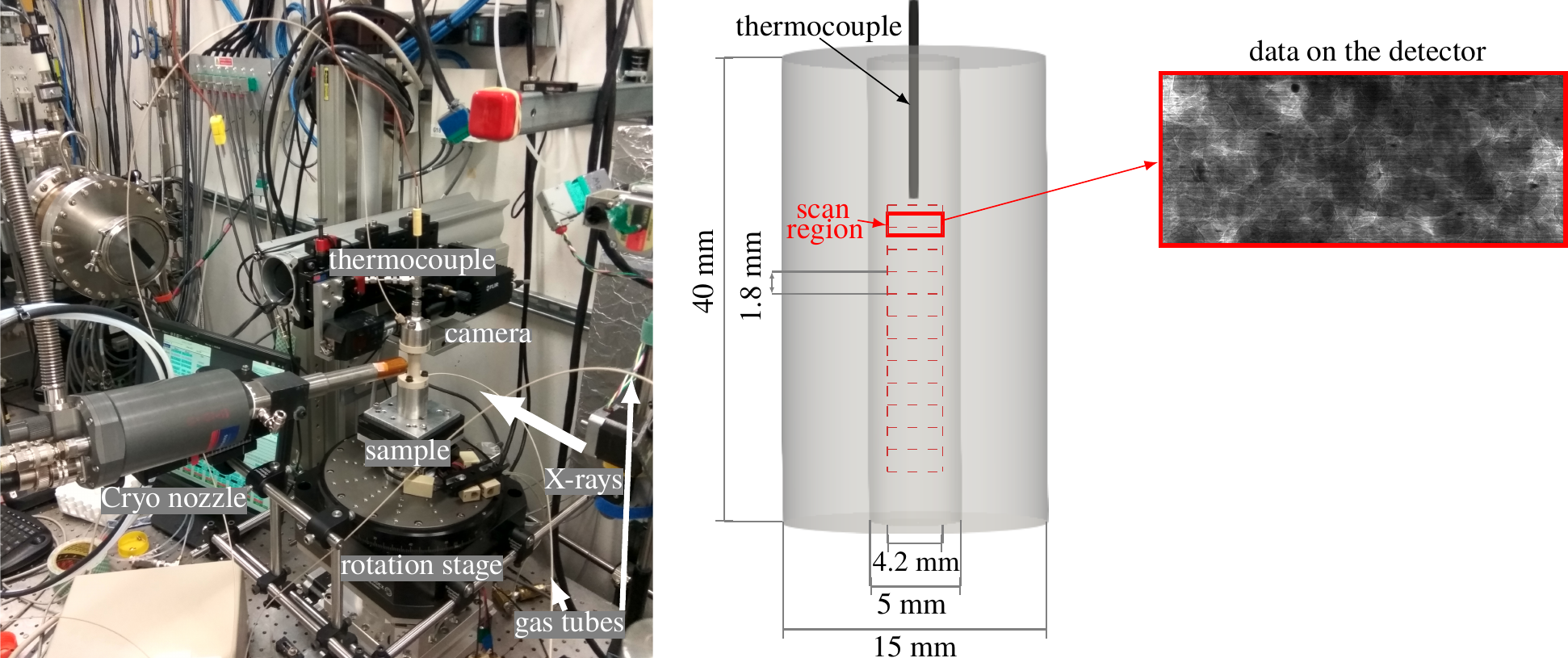}
  \caption{
  %\textcolor{red}{I suggest to remove the flat field image and all reference to it. This does not carry any info for the reader} 
  Setup of the tomography experiments at the 2-BM beamline of APS: general view (left) and a scheme of scanning the sample in the environmental cell (center), single-projection data example (right). Dynamic imaging was performed for the region right under the thermocouple sensor marked by solid-red rectangle frame; dashed-red rectangle frames show scanning positions for obtaining vertical data stacks (several times during the experiment).  %\textcolor{green}{Examples of the data recorded by the detector when the sample is in and out of the field of view are marked by red and blue rectangles, respectively.} 
 % The measurements were performed with \SI{27}{\kilo\electronvolt} X-ray energy and \SI{35}{\milli\second} radiation exposure time. 
 }
  \label{fig:setup}
\end{figure*}

Tomographic experiments were conducted at the bending-magnet 2-BM beamline of the Advanced Photon Source (APS) using the setup shown in Figure~\ref{fig:setup}, left. The environmental cell is the same as the one used in \citep{fusseis2014low} except that the middle vessel is produced with Polyether ether ketone (PEEK) material for better X-ray penetration. The cell is filled with wet sand, and methane gas is served via PEEK high-pressure tubes from the top and bottom sides of the cell. A thermocouple sensor for controlling the temperature in the sample is inserted from the top side. Oxford 700 Cryostream system is used for cooling the sample by a flow of nitrogen gas of low temperature. Methane gas pressure in the cell is controlled by the Teledyne ISCO D-Series single-pump system.
FLIR Oryx ORX-10G-51S5M camera with $2448\!\times\!2448$ pixels (pixel size \SI{3.45}{\micro\meter}$\times$\SI{3.45}{\micro\meter}) was used in a fly scan mode where projections are recorded while the sample is continuously rotated. We were able to measure data from the detector region $2448\!\times\!1024$ because of the reduced size of the X-ray beam passing the monochromator adjusted for the energy of \SI{27}{\kilo\electronvolt}.  The camera recorded projections from a \SI{100}{\micro\meter}-thick LuAG:Ce single-crystal scintillator, magnified through a $2\times$ lens yielding a resulting isometric voxel size of \SI{1.725}{\micro\meter}.

For the experiment we used Ottawa fine white sand (grain sizes 125-\SI{250}{\micro\meter}), chemically pure 2.5-grade methane gas, deionized water, sodium bromide \ch{NaBr} and potassium iodide \ch{KI} (both SIGMA-ALDRICH, purity\!$~>~$\!99\%). The sample was prepared by mixing sand with salty water and packed into the environmental cell. We performed experiments for 10\% mass fraction of water in relation to sand, which corresponds to the excess-gas condition of hydrate formation. We used water with different salinity levels for achieving better  X-ray contrast between water, methane, and gas hydrate. 

A scheme for scanning the environmental cell is shown in the center of Figure~\ref{fig:setup}. The sample is located inside a cylindrical vessel of the height \SI{40}{\milli\meter}, with \SI{5}{\milli\meter} inner and \SI{15}{\milli\meter} outer diameters. %\textcolor{green}{The tube is made from the PEEK material, so its border is almost transparent to X-rays. The scan region used for analysis of the gas-hydrate formation process at short time periods is marked with a red rectangle.- ALREADY WRITTEN ABOVE!! } 
The scanned region is  \SI{4.2}{\milli\meter}$\times$\SI{1.8}{\milli\meter} large and is located close to the thermocouple for more accurate temperature monitoring, see solid-red rectangle frame in Figure~\ref{fig:setup}. During the experiment we also performed several overview scans of the whole sample to analyze water re-distribution in time. The dashed-red rectangle frames in the figure show scan positions for obtaining vertical data stacks by using a total of 12 scan positions.

The right-most part of Figure~\ref{fig:setup} demonstrates an example of recorded data for one projection angle after scanning the sample at the desired region.
%\textcolor{green}{, as well as a flat-field data measured where the sample is out of the field of view and used in the reconstruction process for suppressing artifacts from the detector and scintillator.} 
For the measurements we used a monochromatic X-ray beam of the energy \SI{27}{\kilo\electronvolt} and the exposure time \SI{35}{\milli\second}.  %\textcolor{green}{The parameters were chosen according to the energy range of a high photon flux level at the 2-BM beamline, and according to the optimal photon-counting distribution estimated by scanning samples with similar properties. }
Tomographic projections were collected in a fly scan mode while the sample was continuously rotated over 180$^{\circ}$ at 2.3$^{\circ}$s$^{-1}$, yielding \SI{70}{\second} for each 180$^{\circ}$-scan of 2000 projection angles.  %\textcolor{green}{We also performed  tests with the number of angles satisfying the Nyquist sampling criterion for computed tomography. According to the criterion, high-quality reconstruction can be achieved by choosing the number of angles as $\pi/2$ of the scan width in pixels, which in our case is computed as $\pi/2\cdot2448=3672$. However, we did not observe any significant improvement of reconstruction quality compared to the measurements with 2000 angles.  }

For data processing we use the SAVU tomography framework \citep{wadeson2016savu} consisting of a set of plugins for tomographic data processing and reconstruction. The constructed pipeline for studying gas-hydrate formation processes includes plugins for flat field correction, ring removal, and phase-retrieval filter.  We also prepared our own plugin for tomography data reconstruction by using the log-polar-based imaging algorithm optimized for GPUs \citep{andersson2016fast}.

% In Figure~\ref{fig:rec} we present reconstructions with and without applying the phase-retrieval procedure by using the Paganin filter \citep{paganin2002simultaneous}. Such contrast improvement in phase-retrieved reconstruction is caused by the usage of the monochromatic coherent beam (with fixed X-ray energy \SI{27}{\kilo\electronvolt}) extracted from a high-intensity white beam coming directly off the synchrotron. %This effect cannot be achieved by employing standard laboratory X-ray sources.

Following \citep{lei2018salinityForCT} we conducted experiments for choosing optimal water salinity for optimal contrasts between different phases: sand grains, water, methane hydrate, gas. 
By testing two most popular contrast-enhancing compounds, sodium bromide \ch{NaBr} and potassium iodide \ch{KI}, we picked the first one since it has higher X-ray attenuation for the energy range \SI{20}{}-\SI{30}{\kilo\electronvolt} which is optimal for the 2-BM beamline. Then we conducted two experiments on imaging gas-hydrate formation in sand with water of different salinity levels: 3.5\% \ch{NaBr} water concentrate which is close to that of the natural sea water, and 10\% \ch{NaBr} solution for better contrast between water and hydrate. 
These salinity values are consistent with data for natural hydrate-bearing samples. Pore water salinity for the natural samples from the Mallik 2L-38 well, Mackenzie Delta, Canada, is between 4\% and 12\% \citep{winters2007salt}. And the fluids recovered from the Mount Elbert Well, Alaska North Slope, show salinity values from 2.5\%  to 7.5\% \citep{torres2011salt}. 

\begin{figure*}
  \centering
  \includegraphics[width=0.77\textwidth]{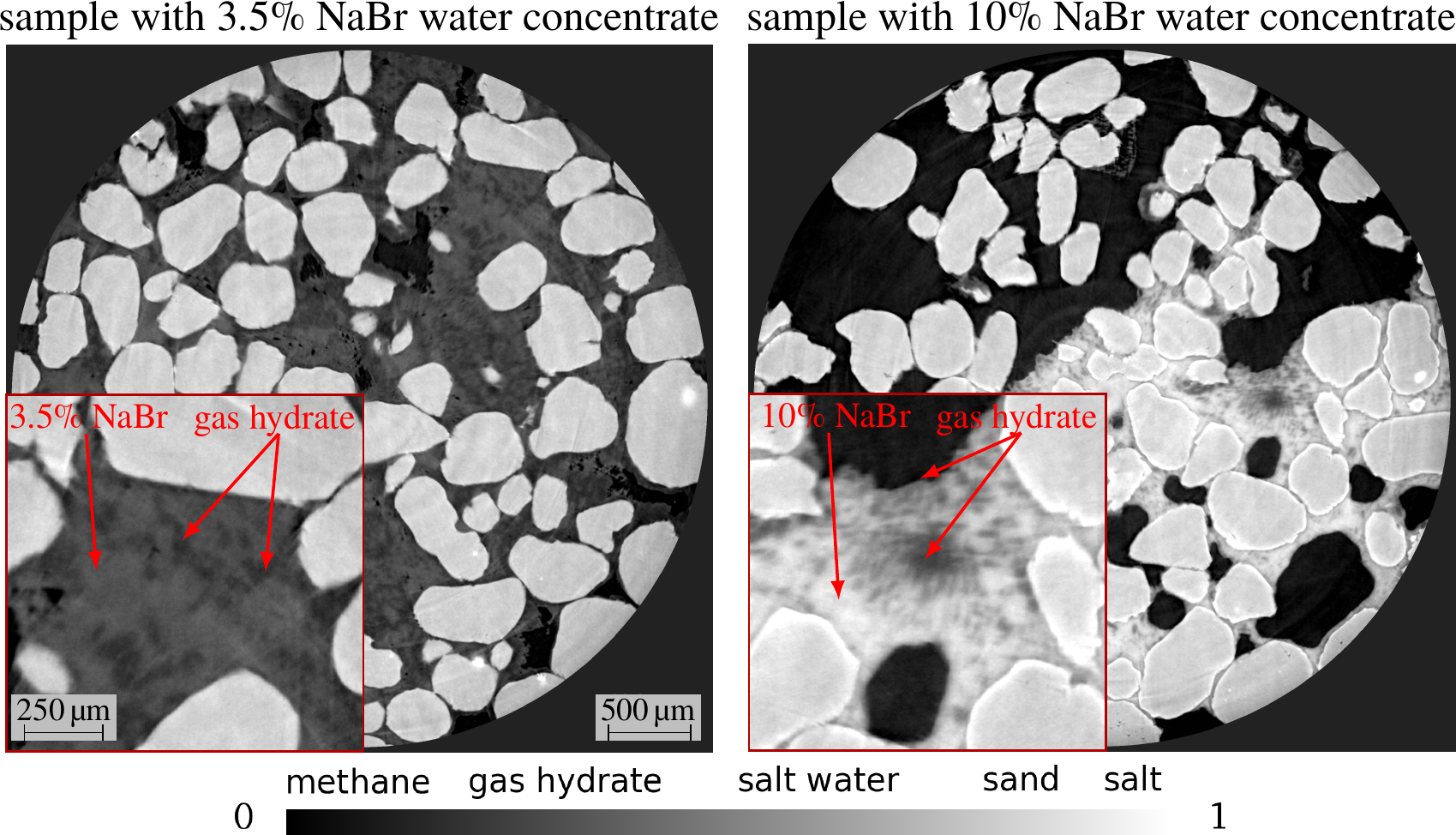}
  \caption{Examples of reconstructed data slices through the samples containing gas hydrates formed in 7 hours with 3.5\% and 10\% \ch{NaBr} water concentrates. The colorbar shows normalized X-ray attenuation of the materials inside the samples.} %The presented slices have dimensions ($2448\!\times\!2448$) and were reconstructed by using the data from 2000 projection angles acquired in \SI{70}{\second}.}
  \label{fig:rec}
\end{figure*}

In Figure ~\ref{fig:rec} one can see examples of reconstructed images during the hydrate formation from
water with salinity 3.5\% (left) and 10\% (right). 
The figure is equipped with the colorbar: black corresponds to the methane gas, dark gray - to the gas hydrate, light gray - to salty water, which in the case of 10\% \ch{NaBr} solution is very close to or even lighter gray of sand grains. White color also appears at the end of the experiments when the hydrate formation consumes almost all water leaving very salty brine or even pure salt.    According to reconstruction results, the contrast between gas hydrate and water for the 3.5\% NaBr is not sufficient to reliable segment out the gas hydrate. Therefore, we chose to use 10\% \ch{NaBr} water concentrate for further experiments. In this case, the attenuation coefficient of salty water becomes close to the coefficient of sand, which is not a problem though because sand particles are almost static during the experiment and can clearly be distinguished from water by using standard segmentation techniques.

%%%%%%%%%%%%%%%%%%%%%%%%%%%%%%%%%%%%%%%%%%%%%%%%%%%%%%%%%%%%%%%%%%%%%%%%%%%%%%
\section{Imaging Results and Discussion}

In this section, we present results by describing different phenomena observed inside the sample during the whole experiment time.
To simulate gas-hydrate growth we followed certain temperature and gas pressure conditions, see \citep{sun2014condition}. Specifically, we used the temperature T~=~\SI{7}{\celsius} and the pressure P~=~\SI{10}{\mega\pascal}, which correspond to the methane-hydrate stability zone. Temperature is positive so that all non-smooth boundaries appearing correspond to the methane gas hydrate (water forms smooth boundaries with curvatures formed by the surface tension forces).  Note that we use the excess-gas method of the hydrate formation in this case as we have about 10~\% initial water saturation in the sand sample. 

We followed the gas-hydrate formation process by scanning the sample automatically every 15 minutes.  We also performed continuous scanning (without \SI{15}{\minute} delays) during the hydrate-decomposition stage. Each full scan is used to reconstruct an image of the sample volume of size $4.2\times4.2\times1.8$ \SI{}{\cubic\milli\meter} ($2448\times2448\times1024$ pixels). At certain times we performed scanning of the whole sample as a vertical stack of 12 scan regions, cf. Figure~\ref{fig:setup}.
For better understanding the processes, we were zooming into the particular pore-space sub-volumes of the size $0.9\times0.9\times0.9$ \SI{}{\cubic\milli\meter} ($512\times512\times512$ pixels) as shown in the following Figures \ref{fig:two_types_gh}-\ref{fig:dissociation}.

%\textcolor{red}{I think that the following paragraph could be more appropriate in the introduction! }
Later in the text we present some observation on the dynamic processes during the methane-hydrate formation. We find it convenient to refer to different gas-hydrate formation types reported from previous micro-CT imaging studies:
\begin{enumerate}
  \setlength\itemsep{0.01cm}
  \item Formation of the hydrate film at the water-gas boundary \citep{lei2019pore}; this hydrate shell should form quickly and remain stable after formation. 
  \item Hydrate growth into the gas pocket by formation of hydrate spikes \citep{lei2019pore}; spike formation should take some time and should be noticed on successive images. 
 \item 
 gas hydrates were considered to nucleate and grow as lens-shaped clusters in the pore space without contact to grains in \citep{zhao2015microstructural}; in the dynamic imaging it should look like growing hydrate clusters.    
\end{enumerate}

Normally, the hydrate shell (type 1) should prevent the exchange of gas and water slowing down and preventing the hydrate formation of type 2 or 3. Thus \citep{chen2018fastXe} proposed various hypotheses explaining Xenon hydrate formation in sandpack in the case when there was no obvious access to free water (thin water film that forms a capillary bridge with spontaneous water recharge or water vapor mass transport through hydrate shells). 
%Water on hydrophobic surfaces can be transported through the gas phase with no connected water films.
Our studies show high mobility of water, i.e. gas-hydrate formation creates structures which do not prevent water flows with no considerable water volumes getting trapped in the hydrate shells.  

%%%%%%%%%%%%%%%%%%%%%%%%%%%%%%%%%%%%%%%%%%%%%%%%%%%%%%%%%%%%%%%%%%%%%%%%%%%%%%
\subsection{Water Migration}

Water migration in the pore space during hydrate formation was first observed by \citep{gupta2006medicalCT} and \citep{kneafsey2007medicalCT}. This observation was confirmed later by using a micro-focus X-ray CT apparatus \citep{yang2015labCT, lei2019flows}. It can be explained by the cryogenic suction phenomenon, i.e. capillary forces are pulling water towards  regions becoming water-depleted due to hydrate formation.  
\cite{lei2019flows} reported overall displacement of water towards the metallic walls of the chamber (that should serve as an efficient heat dissipation sink).

\begin{figure*}%[htbp]
  \centering
  \includegraphics[width=0.9\textwidth]{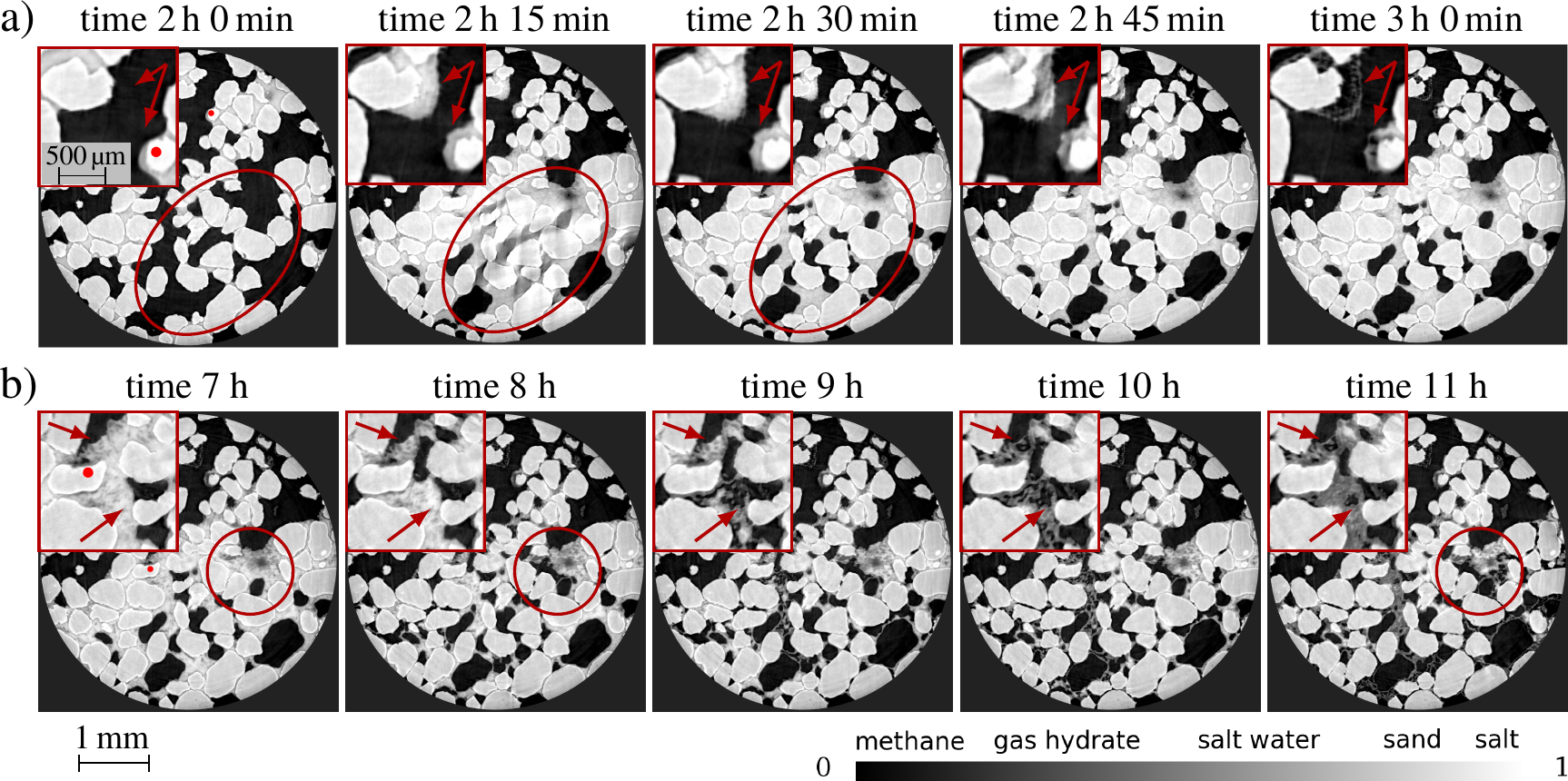}
  \caption{Water migration in the gas-hydrate formation process: a) initial phase of the hydrate formation (upper row) - pores are empty at 2 h, we see imaging artifacts at 2~h~15~min indicating water movement in progress, water settles down at 2~h~30~min supporting the hydrate formation; 
  b) final phase of the hydrate formation (bottom row) - water is mostly leaving this particular slice in a few stages unveiling complicated hydrate structure and supporting continuing hydrate formation
  }
  \label{fig:water_migration}
\end{figure*}

We also observed migration of considerable water volumes. Water flows happened randomly, several times during the experiment. It first occurred 1.5 hours reaching gas-hydrate stability PT-conditions. Then water flows repeated several times in the following 12 hours. Water movement episodes are illustrated in Figure~\ref{fig:water_migration}. The same horizontal slice is shown in both rows. The upper row (Figure~\ref{fig:water_migration}a) shows the initial phase of the hydrate formation. Big red oval highlights typical water migration during this stage. There was mostly gas in pores in these regions 2 hours after the experiment beginning. At 2~h~15~min we see imaging artifacts indicating water movement in progress. It means that most of the water comes in in less than 70 s (scanning time). At 2~h~30~min we see that all the water is settled. 
Even more complicated water migration behaviour is shown in the zoom-in window in the upper row (Figure~\ref{fig:water_migration}a): the pore is empty at 2~h~00~min, water appears on the grains at 2~h~15~min, it starts flowing away at 2~h~45~min producing imaging artifacts, at 3~h~00~min water have moved away leaving gas-hydrate shell marking previous water position. 
At later times water is mostly leaving this particular slice 
in a few stages,
unveiling complicated hydrate structure, see regions marked by red circles in Figure~\ref{fig:water_migration}b (bottom row). In the zoom-in windows one can see a similar process: water stays in the pore at 7 and 8~h, then part of water disappears leaving the hydrate structure at 9~h, hydrate formation continues also attracting some additional water at 11~h.

A few summarizing observations on water migration. First, water flow is not continuous. Water does not move slowly for a long time as one could expect from considering the phase transition process and action of capillary forces. It repeatedly moves several times staying steel in between. Second, water moves fast. Imaging artifacts or blurring (e.g. Figure~\ref{fig:water_migration}, panel~2~h~15~m) indicates that large water volumes move faster than the scanning time of \SI{70}{\second}. Third, hydrate structures remain permeable for water. Water movement is not prevented by the hydrate shells and does not destroy them (e.g. zoom-in window in Figure~\ref{fig:water_migration}, panel~3~h and 10~h). These hydrate structures are similar to those shown in the second figure in \citep{chaouachi2015microstructural}. Note that if successive images are not available then these thin hydrate structures can be misinterpreted as the hydrate growth of type 2 (growth into a gas pore) from \citep{yang2015labCT, lei2019flows}.    

\begin{figure}%[htbp]
  \centering
  \includegraphics[width=0.37\textwidth]{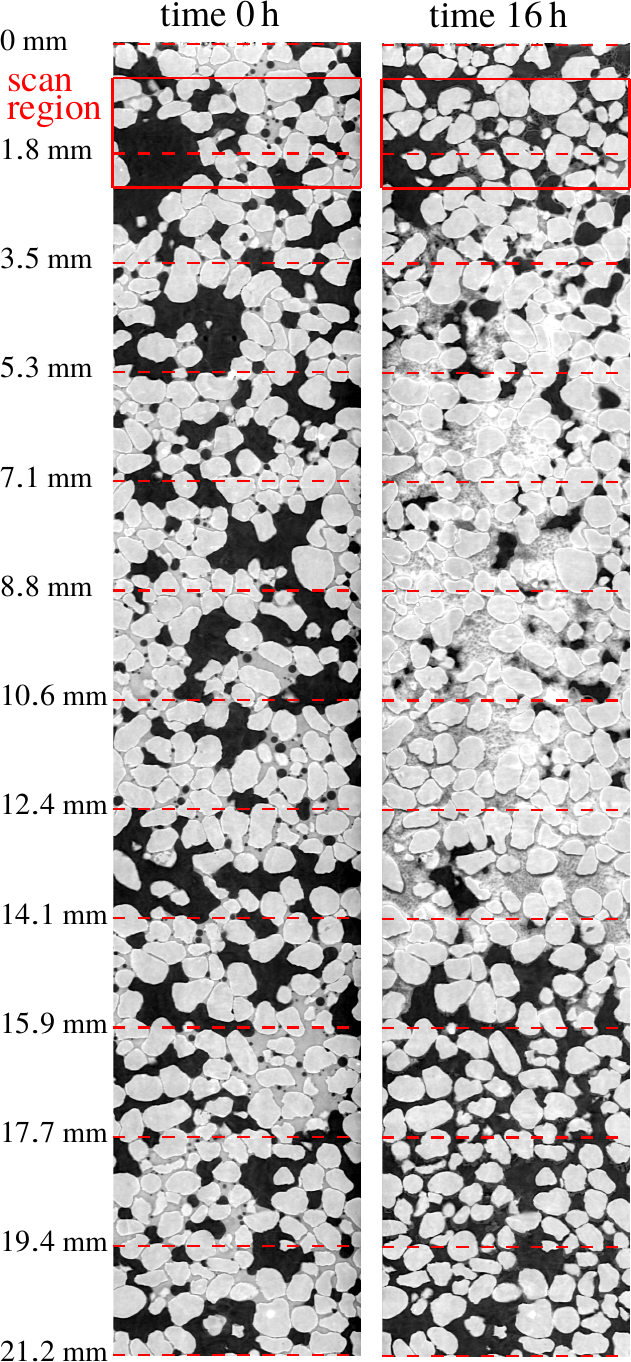}
  \caption{Water/hydrate distribution in the sample: at the beginning of the experiment (left) and after 16 hours of gas-hydrate formation (right). Each vertical stack represents 12 images merged after reconstruction as shown in Figure~\ref{fig:setup}, center. Red rectangle corresponds to the scanning region discussed in details in this section. }
  \label{fig:water_migration_vert}
\end{figure}

The overall pattern of water migration can be understood while comparing vertical sections through the full sample at the beginning of the experiment (Figure~\ref{fig:water_migration_vert},~left) and after a sufficiently long period of gas-hydrate formation (Figure~\ref{fig:water_migration_vert},~right). Overall water moved away from the top and bottom of the sample to support predominant gas-hydrate formation in the central part of the sample.
After hydrate decomposition, water has redistributed in the sample more uniformly again due to capillary forces.
In our experiments we did not notice a tendency of preferable hydrate formation near the sample side walls as reported in \citep{lei2019flows}. The reason could be the difference in the material of the walls of the environment cell: in our case they were made of plastic (PEEK) as opposed to metal walls in \citep{lei2019flows}. 

%%%%%%%%%%%%%%%%%%%%%%%%%%%%%%%%%%%%%%%%%%%%%%%%%%%%%%%%%%%%%%%%%%%%%%%%%%%%%%
\subsection{Different Types of Gas-hydrate Formation}

\begin{figure*}
  \centering
  \includegraphics[width=0.8\textwidth]{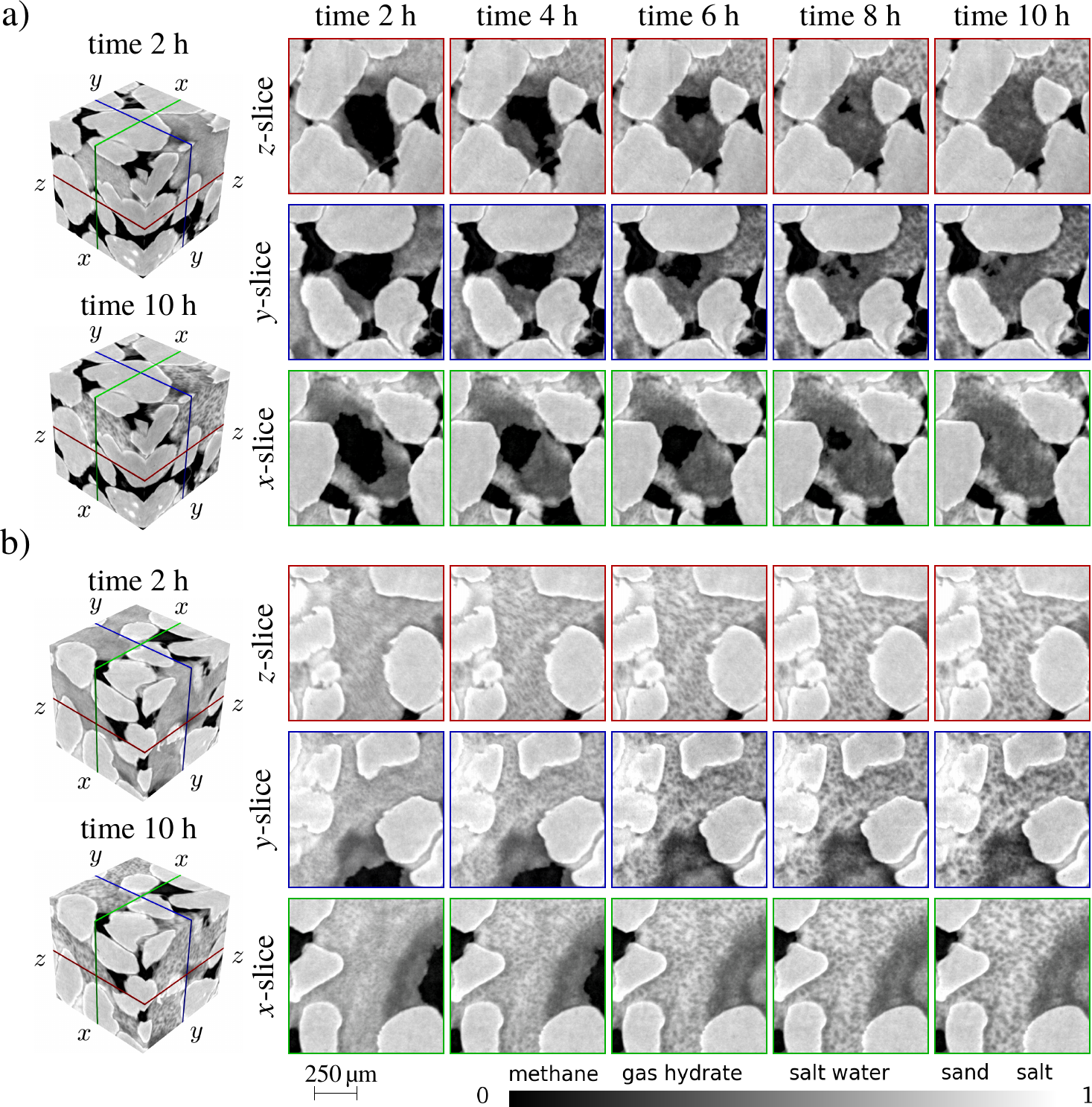}
  \caption{Images of the sample subvolumes illustrating  gas-hydrate growth into gas pockets a), and hydrate formation in water-filled pore b) for different times (\SI{2}{\hour} - \SI{10}{\hour} from the experiment beginning); sand grains are static during the whole period. Left column: 3D zoomed sub-volumes; right: $x$-, $y$- and $z$-slices through these sub-volumes at different times. 
  }
  \label{fig:two_types_gh}
\end{figure*}

In Figure~\ref{fig:two_types_gh} we show two zooms into particular pores of our sample, that illustrate the two types of the hydrate growth: 
\begin{enumerate}
  \setlength\itemsep{0.01cm}
  \item Growing into a gas pocket (central pore in Figure~\ref{fig:two_types_gh}a and right-bottom pore in Figure~\ref{fig:two_types_gh}b). 
  \item Hydrate formation in a water-filled pore (central pore in Figure~\ref{fig:two_types_gh}b). 
\end{enumerate}

The process of growing into the pore is clearly seen in the successive images. Note that the central pore in z-slice gets fully filled with the hydrate in 8 hours while left and right pores in y-slice remain free of the hydrate. It means that continuous water supply goes into one pore and does not go into the other. 
Also note that in this case we do not see the hydrate spikes as reported in \citep{lei2019flows}. Pore filling looks more like a frost growth. 

The process of hydrate formation in a water-filled pore is shown in the central pore in Figure~\ref{fig:two_types_gh}b. One can see how homogeneous gray region gains ``foam'' structure - becomes segmented with time due to formation of hydrate particles (darker regions) and separation of salty water (lighter regions). In some places the salinity of the remaining water gets so high that it becomes even brighter than sand particles. Note again that $x$- and $y$-slices in Figure~\ref{fig:two_types_gh}b show both types of hydrate growth: growing into a pore and formation in a water.
Note that formally we have the excess-water environment for the hydrate formation in this case. However, we see that the hydrate formation rate in this high assuming efficient mechanism of the methane transfer inside the water-filled pore. 

%%%%%%%%%%%%%%%%%%%%%%%%%%%%%%%%%%%%%%%%%%%%%%%%%%%%%%%%%%%%%%%%%%%%%%%%%%%%%%
\subsection{Self-preservation Effect}

\begin{figure*}
  \centering
  \includegraphics[width=0.8\textwidth]{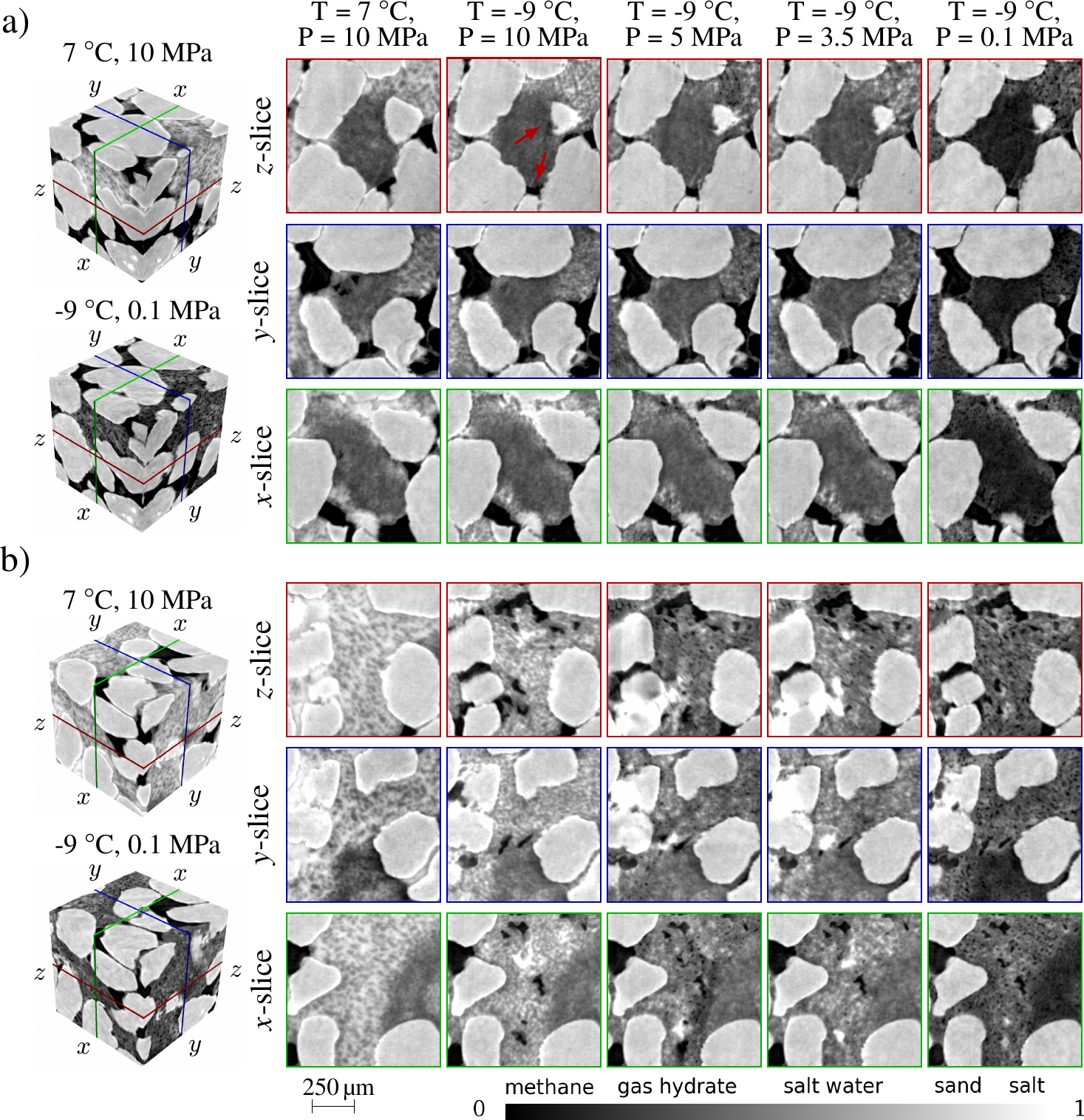}
  \caption{Self-preservation effect of gas hydrates formed into gas pockets a) and in water volumes b) after freezing the sample from the temperature T~=~\SI{7}{\celsius} to T~=~\text{-}\SI{9}{\celsius} and dropping the pressure from P~=~\SI{10}{\mega\pascal} to P~=~\SI{0.1}{\mega\pascal}; sand grains are moved whenever the sample is frozen. Left column: 3D zoomed sub-volumes; right: $x$-, $y$- and $z$-slices through these sub-volumes at different times. 
  }
  \label{fig:preservation_gh}
\end{figure*}

In this experiment, we also check the self-preservation effect of gas hydrates, see \citep{stern2001preservation, hachikubo2011preservation, chuvilin2018preservation}. It is known that gas hydrates can be stored at atmospheric pressure below the melting point of ice, even though this condition is outside of the hydrate stability zone. After freezing the sample to T~=~-\SI{9}{\celsius} we were gradually dropping the gas pressure from P~=~\SI{10}{\mega\pascal} (gas hydrate stability zone) to P~=~\SI{0.1}{\mega\pascal} (atmospheric pressure) and acquire tomography data for different pressure levels. As shown in Figure~\ref{fig:preservation_gh}, the gas hydrate formed into gas pockets (Figure \ref{fig:two_types_gh}a) evaporates whenever the gas pressure dropped to \SI{0.1}{\mega\pascal}. At the same time, the gas hydrate formed in water volumes is preserved. Note that water inflows and outflows during the pressure drops do not affect the hydrate structure. This means that the method of hydrate formation -- excess-gas or excess-water \citep{waite2009physical, manakov2017laboratory}, has a great impact on the self-preservation effect of formed hydrates. And also, hydrate formation type should be taken into account in the modeling of gas production from hydrate deposits located in permafrost regions (for example, in \citep{sun2014production}).

%%%%%%%%%%%%%%%%%%%%%%%%%%%%%%%%%%%%%%%%%%%%%%%%%%%%%%%%%%%%%%%%%%%%%%%%%%%%%%
\subsection{Decomposition}

\begin{figure*}
  \centering
  \includegraphics[width=0.84\textwidth]{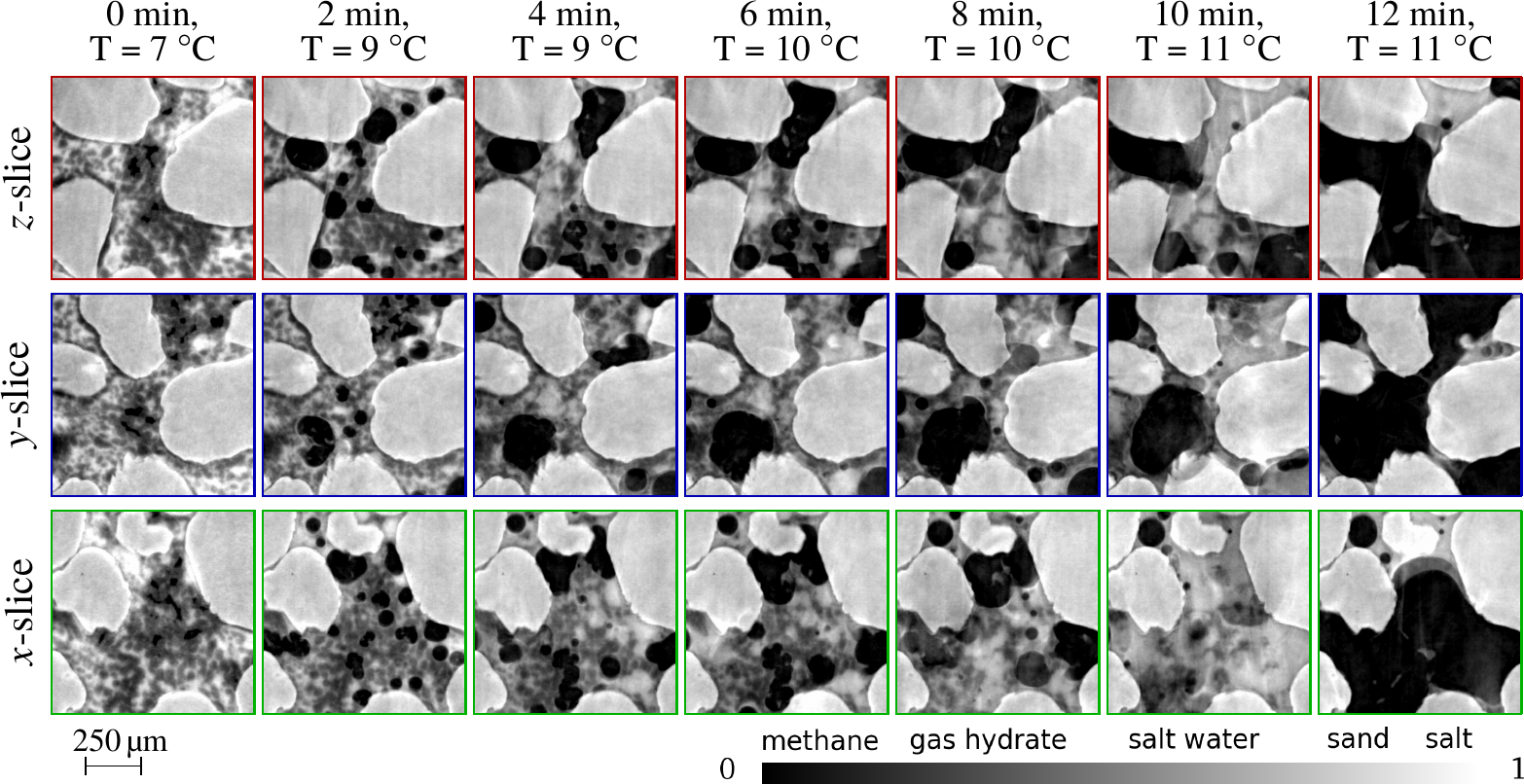}
  \caption{Capturing fast gas-hydrate decomposition process while increasing the temperature from T~=~\SI{7}{\celsius} to T~=~\SI{11}{\celsius}. Slices over $x$-,$y$-, and $z$-axes are extracted from the 3D zoomed sub-volume evolving with the temperature increase.} 
  \label{fig:dissociation}
\end{figure*}

Synchrotron-based tomography allows capturing fast changes of the sample inner structure. The gas-hydrate decomposition process is a fast process even when it is caused by temperature increase (not pressure drop). New volumes of water and gas appear causing their redistribution in the pore space. 
In Figure~\ref{fig:dissociation} we demonstrate reconstructions of the data acquired every 2 minutes. We see seven successive images covering 12 minutes of the temperature growth from T~=~\SI{7}{\celsius} to T~=~\SI{11}{\celsius}. We see the formation of methane bubbles at the beginning. With increasing temperature one can still follow faster movement of gas, water and hydrate particles, even though there are some motion artifacts in the last images. 

In fig.~3 of \citep{yang2016synchrotron} the authors show decomposition of Xenon hydrate with a \SI{20}{\min} interval. First, we see that the hydrates formed from Xenon and methane have different morphology. Xenon hydrates mostly form envelopes attached to the sand grains, whereas methane hydrates have more complex heterogeneous structure. Second, the Xenon hydrate started to dissociate from water-gas boundaries, forming meniscus-shaped water layers. In the methane hydrate case the decomposition process starts in the whole hydrate volume with forming of internal methane bubbles, see Figure~\ref{fig:dissociation}. This comparison confirms the need to use methane instead of Xenon in studies of gas hydrates.

%%%%%%%%%%%%%%%%%%%%%%%%%%%%%%%%%%%%%%%%%%%%%%%%%%%%%%%%%%%%%%%%%%%%%%%%%%%%%%
\subsection{Matrix Deformation}

We noticed an interesting difference between water freezing and the hydrate formation. \citep{lei2019flows} mentioned that matrix deformation (movement of sand grains) were caused by the hydrate formation and freezing processes. In our experiments we see that grains do not move during the hydrate formation, for example, see Figure~\ref{fig:two_types_gh}, top row. However, freezing always caused some grain movement, see Figure~\ref{fig:preservation_gh}, top row. One can see that the image of the triangle-like grain changed between the first image in the row (T~=~\SI{7}{\celsius}, T~=~\SI{10}{MP}a), and the second image in the row (T~=~-\SI{9}{\celsius}, T~=~\SI{10}{MP}a). While checking other images we see that freezing always caused the grain movement while the hydrate formation did not cause the particle displacement. This difference in mechanical interaction between the hydrate particles and the matrix grains may cause a difference in mechanical properties between frozen and hydrate-bearing materials. 

%%%%%%%%%%%%%%%%%%%%%%%%%%%%%%%%%%%%%%%%%%%%%%%%%%%%%%%%%%%%%%%%%%%%%%%%%%%%%%
\section{Conclusions and Outlook}

In this paper, we visualized the formation and decomposition of methane gas hydrates using in-situ X-ray synchrotron tomography at \SI{1.7}{\micro\meter} voxel size in 3D. The excellent spatio-temporal resolution of the instrument (single scan time was \SI{70}{\second}) allows for detailed insight into the processes of interest with the added value of sufficient statistics (over 500 3D volumes were acquired over the course of the experiment) 

%\textcolor{green}{For our three-day tomographic experiment at the synchrotron we acquired about 500 datasets of more than 2.3Tb size in total. Fast scanning times (70 seconds) provided a lot of successive images giving a detailed insight into the processes associated with the methane-hydrate formation/decomposition. - REFORMULATE ABOVE }

We experimentally observed the time delay between formation of favorable PT-conditions and the actual start of the methane-hydrate formation. It took about 2 hours in the first experiment, and about 1.5 hours in the second one. We associate the beginning of the hydrate formation with the first water movement observed. Our studies show high mobility of water, i.e. gas-hydrate formation forms structures which do not prevent water flows with no considerable water volumes getting trapped in the hydrate shells. 

During the hydrate formation we observed water movements that we associate with the cryogenic suction (capillary forces due to water consumption into the hydrate formation). 
It was not a steady slow flow of water but several fast movements (less than \SI{70}{\second} in duration) of considerable water volumes followed by steady-state periods. These fast water redistribution events happened in different areas at different times most likely controlled by the hydrate formation dynamics. In our experiments we did not notice a tendency of preferable hydrate formation near the sample side boundaries as reported in \citep{lei2019flows}, in contrary gas hydrates formed preferably in the central regions. A possible explanation is that the walls of high-pressure chambers were made out of different materials: plastic in our case, metal in \citep{lei2019flows}. 
%\textcolor{red}{the following statement I cannot really see on figure 4, or I do not fully understand the statement} 
We also observed that after the hydrate decomposition, water has redistributed to form the more uniform distribution. This effect of water redistribution should be studied in more detail and should be taken into account during the laboratory studies of physical properties of hydrate-bearing samples. 

During the hydrate formation we captured two types of the gas-hydrate growth: into gas pockets and inside water volumes. These two gas-hydrate types show different properties during the decomposition in the self-preservation mode - pressure drop at negative temperatures. 
Gas hydrate formed in water volumes remains more stable compared to hydrate formed as growing into the gas pocket.
This means that the history of the gas-hydrate formation influences its behaviour at the decomposition stage (e.g gas-hydrate production).

We noticed an interesting difference between two crystallization processes: water freezing and hydrate formation. In our experiments the water freezing caused the particle pushing (displacement of the sand grains), the hydrate formation did not result in the particle displacement. This difference in mechanical interaction between the hydrate particles and the matrix grains may cause a difference in mechanical properties of frozen and hydrate-bearing geomaterials. 

Our results confirm a great variety of fast processes associated with methane gas-hydrate formation-decomposition. This justifies the necessity of using fast continuous scanning of the sample with many non-delayed rotations in order to capture fast water, gas, and gas-hydrate redistribution processes. In \citep{nikitin2019four} we introduced a method for suppressing motion artifacts in the data acquired with continuous scanning and tested it on real data sets. Combining the developed method and fast acquisition we envision a 10-fold decrease of the scanning time that will help in suppressing motion artifacts observed in some of the images presented here.

%%%%%%%%%%%%%%%%%%%%%%%%%%%%%%%%%%%%%%%%%%%%%%%%%%%%%%%%%%%%%%%%%%%%%%%%%%%%%%
\section*{Acknowledgments}

This research used resources of the Advanced Photon Source, a U.S. Department of Energy (DOE) Office of Science User Facility operated for the DOE Office of Science by Argonne National Laboratory under Contract No. DE-AC02-06CH11357. 
The work is supported by the Swedish Research Council grant (2017-00583). The authors are grateful to A.D. Duchkov, A.Yu. Manakov, and K.E. Kuper for helpful discussions. 

%%%%%%%%%%%%%%%%%%%%%%%%%%%%%%%%%%%%%%%%%%%%%%%%%%%%%%%%%%%%%%%%%%%%%%%%%%%%%%

\section*{References}
\bibliography{refs}

\end{document}